\input{aipcheck}

\documentclass[
    ,final            
  ,numberedheadings 
  ]
  {aipproc}

\layoutstyle{6x9}

\begin{document}

\title{Thermal instabilities in Active Galactic Nuclei: \\ 
the case of thin vs. thick ionized media}

\classification{95.85.Nv -- 95.75.-z -- 52.25.Os}
\keywords      {X-ray -- Computer modeling and simulation -- Emission, 
absorption, and scattering of electromagnetic radiation }

\author{A. C. Gon\c{c}alves}{
  address={LUTH, Observatoire de Paris-Meudon, 5 Place Jules Janssen, 
F-92195 Meudon, France}, altaddress={CAAUL, Observat\'orio Astron\'omico 
de Lisboa, Tapada da Ajuda, P-1349-018 Lisboa, Portugal} 
}

\author{S. Collin}{
address={LUTH, Observatoire de Paris-Meudon, 5 Place Jules Janssen, 
F-92195 Meudon, France }}

\author{A.-M. Dumont}
{address={LUTH, Observatoire de Paris-Meudon, 5 Place Jules Janssen, 
F-92195 Meudon, France }}

\begin{abstract}
 We have studied the thermal instabilities in the context of 
Active Galactic Nuclei (AGN), addressing the cases of 
thin and thick (stratified) X-ray illuminated media. For that, we have  
compared the behaviour of different models in pressure 
equilibrium. Ionized gas in pressure equilibrium show temperature 
discontinuities which, however difficult to tackle, can be modelled 
with the transfer-photoionization code TITAN. Our code allows to 
choose between the hot and the cold stable 
solutions in the multi-branch regime of the S-curve, and then to 
compute a fully consistent photoionisation model with the chosen 
solution. For the first time, it is now possible to compare the 
true stable solution models with the approximate solution model 
normally used.  Our studies may be applied to media in any pressure 
equilibrium conditions, e.g. constant gas pressure, constant total 
pressure, or hydrostatic pressure equilibrium; they can be used 
to model irradiated accretion discs in AGN, the Warm Absorber 
in type 1 AGN, or the X-ray line-emitting gas in type 2 AGN.  
\end{abstract}

\maketitle


\section{Thermal instabilities in thin and thick media} 
First, let us introduce the subject of thermal instabilities in 
thin and thick ionized media, and address the main differences observed 
in each case. It is well known that a photoionized gas in thermal
equilibrium can display a thermal instability (e.g. Krolik et al. 1981). 
The phenomenon manifests itself in the S-shape of the net 
cooling function, or, which is equivalent, through the 
curve giving the temperature versus the radiation-to-gas-pressure
ratio. Such an S-shape curve allows for the co-existence of gas 
at different temperatures and densities for the same pressure ratio.  
For a given value of the radiation-to-gas-pressure ratio, the gas 
can thus be in three states of thermal equilibrium ---   
one of these states is thermally unstable; the two 
others (a ``hot'' and a ``cold'' solution) are stable. 
Such thermal instabilities are, for instance, at the origin 
of the two-phase model for the interstellar medium (Field et al. 1969)  
and the two-phase medium interpretation of the 
Broad Line Region in Active Galactic Nuclei (AGN) (Krolik et al. 1981). 
 
Since Krolik et al., many studies have been devoted to the S-shape 
curve, but always in the optically thin case. Thanks to the most 
recent developments of the TITAN code, described in Gon\c{c}alves et
al. (2007), it is now possible to address the thermal 
instabilities in the case of optically thick, stratified media (note 
that by ``optically thick", we here mean media optically 
thick to the photoionization continuum; in such case, the continuum 
displays absorptions at one or more ionization edges). 

Our modeling can be applied to the  study of the Warm 
Absorber (WA) in type 1 AGN, the X-ray line-emitting gas 
in type 2 AGN, or still the irradiated accretion discs in 
AGN and X-ray binaries.

\begin{figure}
  \centering
  \includegraphics[width=7.25cm,angle=-90]{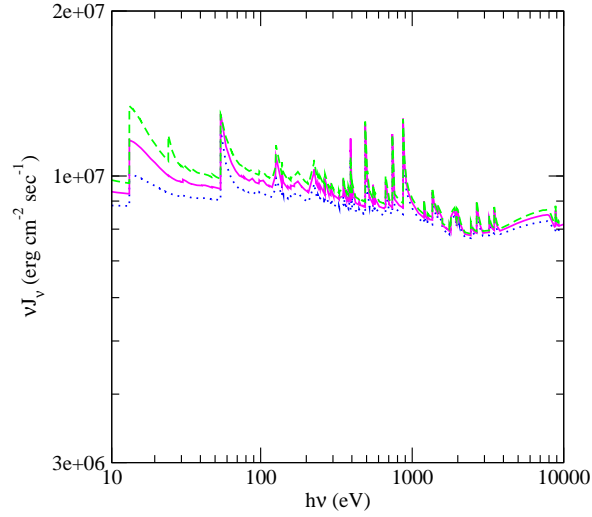}
      \caption{Spectral distribution of the 
mean continuum intensity, computed at the illuminated surface of 
the slab, for three  models   
with the same ionization parameter ($\xi = 1000$) 
and different column densities: $N_{\rm H} = 2\,10^{23}$~cm$^{-2}$ 
(dotted line),  2.5\,10$^{23}$ (filled line), and  3\,10$^{23}$ 
(dashed line). 
For the sake of clarity, the spectral lines have been supressed from 
this figure, where only the continuum is shown; this displays  
important discontinuities, which increase with the slab 
thickness. }
   \end{figure}

\begin{figure}
  \centering
  \includegraphics[width=14.5cm,angle=0]{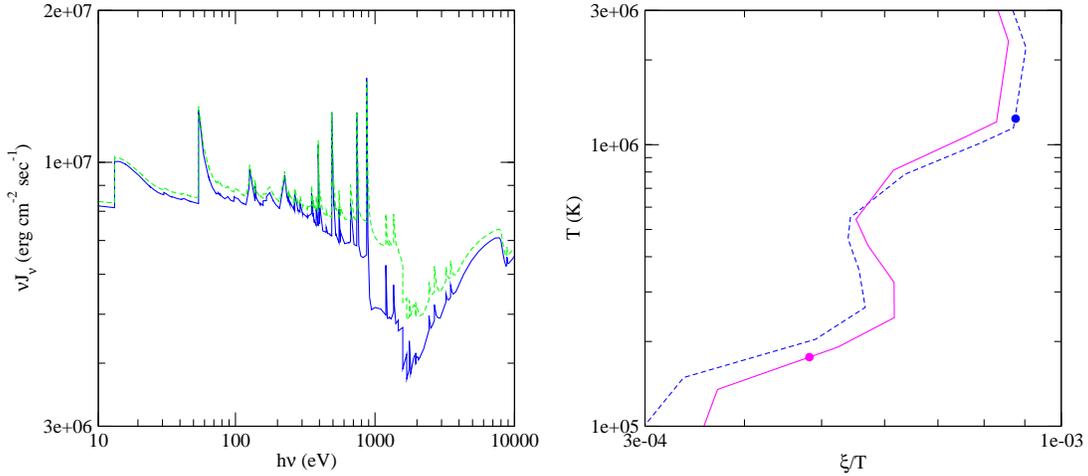}
      \caption{Results for the $\xi = 1000$ and 
$N_{\rm H} = 2\,10^{23}$ cm$^{-2}$ model. Left-hand panel: 
the spectral distribution of the mean continuum 
for two layers located at different depths in the gas slab: 
1.66\,10$^{17}$~cm (dashed line) and 1.90\,10$^{17}$~cm (solid line). 
Right-hand panel: the S-curves giving the temperature 
$T$ versus ~$\xi/T$ for the same two layers;  
the dots represent the equilibrium temperature found for 
each layer.  }
   \end{figure}

\subsection{Optically thin ionized media}
In optically thin media, the radiation pressure keeps a constant 
value, and the energy balance equation solution depends on the 
gas pressure only. When the gas pressure is small compared to the 
radiation pressure, there is a unique stable hot solution, where 
both the heating and the cooling are dominated by Compton processes 
(Compton heating, inverse Compton cooling). As the gas pressure 
increases, atomic processes (photoionization heating,  line and 
continuum cooling) become important, and multiple solutions arise.  
Then, above a given gas pressure, again 
a unique, cold solution, exists.               
Krolik et al. (1981) have studied the thermal instabilities in 
the context of optically thin media. They showed that multiple 
solutions may lead to the existence of a 2-phase medium, with a 
hot dilute medium confining a dense cold one; what about thick media?

\subsection{Optically thick ionized media}
The thermal instability problem in thick, stratified media 
is very different from what happens in thin media, this for 
mainly two reasons:

{\it (i)}~~First, the spectral distribution of 
the mean intensity, $J_{\nu}$, at the illuminated surface 
of the medium is different from the incident spectrum, as 
it equally contains a ``returning" radiation component  
emitted by the slab of gas itself. To illustrate this 
phenomenon, we have computed a series of photoionization 
models with the TITAN code: all models were 
computed under total pressure equilibrium, null 
turbulent velocity, and cosmic abundances;   
the hydrogen numerical density at the illuminated surface, 
$n_{\rm H}$, was  set to $10^{7}$~cm$^{-3}$. We recall that 
the thermal and ionization structures of our ionized gas 
are mainly determined by the ionization parameter $\xi$ 
and the shape of the incident continuum --- 
here a power-law of photon index $\Gamma=2$, covering 
the 10$-$10$^{5}$~eV energy range. Figure~1 shows the 
spectral distribution of the mean continuum intensity at 
the slab surface  for three models with  different values 
of the column density, $N_{\rm H}$; we observe that the 
spectrum  at the irradiated surface  contains strong 
discontinuities, whose amplitude increases  with the slab 
thickness, as does the intensity of the whole spectrum.  

{\it (ii)}~~Second, the spectral distribution 
changes as the radiation progresses inside the medium; 
as a consequence, the shape of the S-curve also changes, 
and instead of traveling along a given S-curve, as the  
radiation-to-gas pressure ratio decreases, the temperature 
follows successively different curves. This behaviour 
is illustrated in Fig.~2, which 
shows the spectral distribution of the mean continuum for two 
layers located at different depths in the gas slab, 
and the corresponding curves giving $T$ versus $\xi/T$. 
We can see that the S-curves are different for the 
two represented layers, and that the deeper 
layer displays a larger multi-branch 
region; this is due to its larger absorption trough.

\section{Summary of our results and their implications}
An important result to keep in mind is that {\it a 
thick, stratified medium, ionized by X-rays, behaves 
differently from a thin ionized medium.}   
This has observational implications in the emitted/absorbed 
spectra, ionization states, and variability. 

It is impossible to know what solution the plasma 
will adopt when attaining the multi-solutions regime:  
it can  oscillate between the hot and cold  solutions, 
it can fragment into hot and cold clumps which will 
coexist together, or it can take the form of 
a hot, dilute medium confining cold, denser clumps. 
In addition, the relative proportion of those phases 
could be varying with time. Nevertheless, one expects 
the  emitted/absorbed spectrum  to be intermediate 
between those resulting from pure cold and hot models. 

TITAN can compute models using either the stable, 
cold/hot solutions (this version of our program is 
based on an isobaric iteration scheme), or else an 
intermediate solution between those two (the computations 
are then based on an isodensity iteration scheme). 
When comparing the results obtained with models based 
on the stable hot/cold solutions,  and the approximate, 
intermediate solution, we conclude that  the hot/cold 
models represent two extreme results corresponding 
to a given gas composition and photoionizing flux. 
We stress that the three (hot, cold, and intermediate) 
models differ not only in the layers where multiple 
solutions are possible, but {\it all along the gas slab}, 
this because the entire radiation field suffers 
modifications while crossing a thick medium. 
The spectra emitted or absorbed by a given ionized medium, 
consisting of a mixture of gas in the hot and cold phases,  
should thus be intermediate between those resulting from 
the pure cold and hot models; therefore, the intermediate 
model provides a good description of such a mixed-phase medium. 
Furthermore, the differences 
between the emitted and absorbed spectra obtained with 
the stable solutions can provide an indication for the 
maximum  error bars  associated to the spectra computed with 
the intermediate solution previously used by TITAN to 
circumvent the problem of thermal instabilities. 
The differences amount at most to 1--2\%\ for the outward 
emitted spectrum, and to 20\%\ for the absorption spectrum. 
The agreement is much better for the outward emission in the 
X-ray range. The lines have similar intensities in all the 
spectra. A more throughout comparison of the  hot, cold, and 
intermediate models is given in Gon\c{c}alves et al. (2007). 

An important point to take into account when choosing 
which computational method to apply, is that the full 
computation of the hot and cold models 
is extremely time-consuming; this is because the process 
is strongly unstable and requires thus more iterations 
than the intermediate model. 
Given the results, we sustain that 
it is reasonable to use the simpler, isodensity scheme to compute 
constant pressure models or hydrostatic equilibrium models.  
This procedure is less {\it ad-hoc} than to choose arbitrarily 
between one of the possible solutions, as other codes do,  
resulting, in the end, in gas structures and emitted/absorbed 
spectra very close to what is expected from an ionized medium 
consisting in similar proportions of gas in the hot and cold phases.

\begin{theacknowledgments}
ACG acknowledges support from the 
FCT, Portugal, under grant ref. BPD/11641/2002.
\end{theacknowledgments}



\bibliographystyle{aipprocl} 




\end{document}